\documentstyle[aps,epsf,multicol,graphicx]{revtex}
\begin{document}
\title{Scaling, Multiscaling, and Nontrivial Exponents 
in Inelastic Collision Processes} 
\author{E.~Ben-Naim$^1$ and P.~L.~Krapivsky$^2$}
\address{$^1$Theoretical Division and Center for Nonlinear Studies,
Los Alamos National Laboratory, Los Alamos, New Mexico 87545}
\address{$^2$Center for Polymer Studies and Department of Physics,
Boston University, Boston, Massachusetts 02215} \maketitle
\begin{abstract}
We investigate velocity statistics of homogeneous inelastic gases
using the Boltzmann equation.  Employing an approximate uniform
collision rate, we obtain analytic results valid in arbitrary
dimension. In the freely evolving case, the velocity distribution is
characterized by an algebraic large velocity tail, $P(v,t)\sim
v^{-\sigma}$. The exponent $\sigma(d,\epsilon)$, a nontrivial root of
an integral equation, varies continuously with the spatial dimension,
$d$, and the dissipation coefficient, $\epsilon$. Although the
velocity distribution follows a scaling form, its moments exhibit
multiscaling asymptotic behavior. Furthermore, the velocity
autocorrelation function decays algebraically with time, $A(t)=\langle
{\bf v}(0)\cdot{\bf v}(t)\rangle\sim t^{-\alpha}$, with a
non-universal dissipation-dependent exponent $\alpha=1/\epsilon$.  In
the forced case, the steady state Fourier transform is obtained via a
cumulant expansion. Even in this case, velocity correlations develop
and the velocity distribution is non-Maxwellian.
\\
\vspace{5pt}
\end{abstract}

\begin{multicols}{2}

\section{Introduction}

Inelastic gases consist of hard sphere particles that interact via
contact interactions and dissipate kinetic energy upon collisions
\cite{pl}. They are used extensively to study dynamics of granular
materials. Numerically, molecular dynamics simulations are quite
successful in modeling many of the observed collective phenomena that
include size segregation, phase transitions, shocks, clustering, and
development of other spatial structures
\cite{h,bssms,gzb,kwg,lcdkg,ou,nbc,slk,rbss}.  In parallel, kinetic theory
is utilized to systematically derive macroscopic properties from the
microscopic collision dynamics \cite{jr,sg,gs}. 

Inelastic gases, a prototype nonequilibrium interacting particle
system, are interesting on their own
rights\cite{pkh,cpy,bm,my,dlk,zk,ep,db}.  Recent theoretical and
experimental studies show that the velocity distributions exhibit
anomalous large-velocity statistics with exponential, stretched
exponential, and Gaussian
tails\cite{lcdkg,ou,nbc,van,rm,ao,italy}. Inelastic gases involve
significant velocity and spatial correlations in contrast with
traditional molecular gases \cite{classical,soto,bbrtv}. Kinetic
theory assumes that spatial velocity correlations are small. While
this assumption can be justified for strongly driven gases, the
situation for freely evolving gases is more difficult since velocity
correlations can be ignored only in the early homogeneous phase
\cite{gz,be,bcdr}, but must be taken into account in the asymptotic
clustering phase.  Clearly, the strong energy dissipation raises
challenging new questions \cite{g}.

Yet, even more elementary questions remain unanswered. For example,
random collision processes effectively generate thermal, purely
Maxwellian, velocity distributions when the collisions are elastic.
In particular, different components of the velocity become
uncorrelated. In this study, we consider these very same processes but
with inelastic collisions.  We show that energy dissipation
fundamentally alters the behavior. The system is intrinsically a
nonequilibrium one, and the resulting velocity distributions are far
from thermal.

We consider a collision process where random pairs of particles
undergo inelastic collisions with a random impact direction.  This
process, often called the Maxwell model, is described by a Boltzmann
equation with a uniform collision rate.  In classical kinetic theory
of gases, the analytically tractable Maxwell model precedes the
Boltzmann equation \cite{max}. Historically, it played an important
role in the development of kinetic theory \cite{true,e,bob}, and it
still remains the subject of current research \cite{bobyl,hayak}.

Very recently, it has been noted that the Maxwell model is
analytically tractable even for inelastic collisions
\cite{bk,bmp,kb,eb,mp}. Interesting behavior emerges in the freely
evolving case. In one dimension, while moments of the velocity
distribution exhibit multiscaling \cite{bk}, the velocity distribution
itself still approaches a scaling form with an algebraic large
velocity tail \cite{bmp}. Here, we show analytically that in arbitrary
spatial dimension the velocity distribution admits a scaling solution
with an algebraic large velocity tail.  The corresponding exponent, a
root of a transcendental equation, depends on the spatial dimension
and the restitution coefficient.  Additionally, we find that the
multiscaling behavior extends to higher dimensions, and that the
velocity autocorrelation function exhibits aging and nonuniversal
asymptotic behavior.  In general, velocity components develop
significant correlations.  Such correlations diminish in the forced
case, although the velocity distribution remains non-Maxwellian.

The rest of this paper is organized as follows.  The basic Boltzmann equation
for the velocity distribution and its Fourier transform are presented in
Sec.~II. In Sec.~III, we investigate the scaling regime, and obtain the
extremal velocity statistics, moments of the velocity scaling function, and
velocity correlations. In Sec.~IV, we illuminate the nonequilibrium dynamics
by studying the time dependent behavior of the moments and the velocity
autocorrelation function. In Sec.~V, we consider nonequilibrium steady states
in the driven case, and obtain the steady state distribution as a cumulant
expansion. A few generalizations are briefly mentioned in Sec.~VI, and 
conclusions are given in Sec.~VII.

\section{The Maxwell Model}

We study a homogeneous system of identical inelastic spherical
particles. The mass and the cross-section are set to unity without
loss of generality. Particles interact via binary collisions that lead
to exchange of momentum along the impact direction. The post-collision
velocities ${\bf v}_{1,2}$ are given by a linear combination of the
pre-collision velocities ${\bf u}_{1,2}$,
\begin{equation}
\label{inel}
{\bf v}_{1,2}={\bf u}_{1,2}\mp(1-\epsilon)\,({\bf g}\cdot {\bf n})\,{\bf n}.
\end{equation}
Here ${\bf g}={\bf u}_1-{\bf u}_2$ is the relative velocity and ${\bf
n}$ the unit vector connecting the particles' centers. In each
collision, the normal component of the relative velocity is reduced by
the restitution coefficient $r=1-2\epsilon$. The energy dissipation
equals $\Delta E = -\epsilon(1-\epsilon)({\bf g}\cdot{\bf n})^2$, so
for $\epsilon=0$ collisions are elastic, while for $\epsilon=1/2$
collisions are perfectly inelastic with maximal energy
dissipation. Since the collision rule (\ref{inel}) is Galilean
invariant, the average velocity can be set to zero without loss of
generality. 

We investigate the ``Maxwell model'' where the collision rate in
the Boltzmann equation equals the typical velocity, rather than
the actual relative velocity \cite{e,bob}. This kinetic theory
describes a stochastic process where randomly chosen pairs of
particles undergo inelastic collisions according to (\ref{inel})
with a randomly chosen impact direction $\bf {n}$. In such a
process,  no spatial correlations develop, and the normalized
velocity distribution function, $P({\bf v},t)$, obeys
\begin{eqnarray}
\label{be} 
{\partial P({\bf v},t)\over \partial t}&=&g \int d{\bf
n} \int d{\bf u}_1\,P({\bf u}_1,t) \int d{\bf u}_2\,P({\bf u}_2,t)\\
&\times &\big\{\delta\left[{\bf v}-{\bf u}_1 +(1-\epsilon)({\bf g}\cdot
{\bf n}){\bf n}\right]-\delta({\bf v}-{\bf u}_1)\big\}.\nonumber
\end{eqnarray}
The overall collision rate equals $g=\sqrt{T}$ where $T$ is the
granular temperature, or the average velocity fluctuation per
degree of freedom, $T={1\over d}\int d{\bf v}\, v^2 P({\bf v},t)$
with $v\equiv|{\bf v}|$. The restriction ${\bf g}\cdot {\bf n}>0$
on the angular integration range in Eq.~(\ref{be}) can be tacitly
ignored, because the integrand obeys the reflection symmetry ${\bf
n}\to -{\bf n}$. This angular integration should be normalized,
$\int d{\bf n}=1$.

We study primarily the freely evolving case where in the absence
of energy input the system ``cools'' indefinitely. From the
Boltzmann equation (\ref{be}), the temperature rate equation is
\begin{equation}
\label{tl} 
{d\over dt}T=-\lambda\,T^{3/2},\qquad{\rm with}\qquad
\lambda={2\epsilon(1-\epsilon)\over d}.
\end{equation}
The constant $\lambda=2\epsilon(1-\epsilon)\int d{\bf n}\, n_1^2$, is
obtained using the identity $n_1^2+\ldots+n_d^2=1$ that yields $\int d{\bf
  n}\, n_1^2=1/d$.  Solving Eq.~(\ref{tl}) we find that the temperature
decays according to Haff's cooling law\cite{pkh}
\begin{equation}
\label{temp} 
T(t)=T_0\,(1+t/t_*)^{-2},
\end{equation}
with the time scale $t_*=d/\left[\epsilon(1-\epsilon)\sqrt{T_0}\right]$ set
by the initial temperature, $T_0$. 

Given the convolution structure of the Boltzmann equation
(\ref{be}), we introduce the Fourier transform \cite{bob} of the
velocity distribution function,
\begin{equation} 
\label{Fourier} F({\bf k},t)=\int d{\bf v}\,e^{i{\bf k}\cdot {\bf
v}}\,P({\bf v},t).
\end{equation}
We conveniently reset the collision rate to unity by modifying the
time variable. The collision counter $\tau$ is defined via the
transformation ${d\over d\tau}={1\over \sqrt{T}}{d\over dt}$. 
Specifically,  
\begin{equation}
\label{tau} \tau={2\over \lambda}\,\ln\left(1+t/t_*\right) 
\end{equation}
equals the average number of collisions experienced by a particle.
Applying the Fourier transform to Eq.~(\ref{be}) and integrating over
the velocities gives
\begin{equation}
\label{fkt} {\partial \over \partial \tau}F({\bf k},\tau)+F({\bf
k},\tau)= \int d{\bf n}\,F\left[{\bf k}-{\bf
q},\tau\right]F\left[{\bf q},\tau\right],
\end{equation}
with ${\bf q}=(1-\epsilon)({\bf k}\cdot {\bf n})\,{\bf n}$.  This
equation reflects the momentum transfer occurring during collisions.

We restrict our attention to isotropic situations, and write the
Fourier transform $F({\bf k },\tau)\equiv F(z,\tau)$ in terms of
the variable $z=k^2$. To perform the angular integration, it
proves useful to employ spherical coordinates with the polar axis
parallel to ${\bf k}$, so that $\hat{\bf k}\cdot {\bf n}=\cos
\theta$. The $\theta$-dependent factor of the measure $d{\bf n}$
is proportional to $(\sin \theta)^{d-2} d\theta$. In terms of the
variable $\mu=\cos^2\theta$ one has $d{\bf n}\equiv{\cal D}\mu$,
with
\begin{equation}
\label{dmu}
B\left({1\over 2},{d-1\over 2}\right){\cal D}\mu=\mu^{-{1\over
2}}(1-\mu)^{d-3\over 2} d\mu
\end{equation}
where $B(a,b)$ is the beta function. This integration measure is
properly normalized, $\int_0^1{\cal D}\mu=1$. Hereinafter, we
denote angular
  integration with brackets
\begin{equation}
\langle \,f \,\rangle=\int_0^1 {\cal D}\mu\, f(\mu).
\end{equation}
The governing equation (\ref{fkt}) for the Fourier transform can
now be rewritten in the convenient from
\begin{equation}
\label{fkt1} {\partial \over \partial \tau}F(z,\tau)+F(z,\tau)=
\big\langle \,F(\xi z,\tau)\,F(\eta z,\tau)\,\big\rangle,
\end{equation}
with the shorthand notations $\xi=1-(1-\epsilon^2)\mu$ and
$\eta=(1-\epsilon)^2\mu$. Hence, the Fourier equation is both
non-linear and non-local. Interestingly, while it is difficult to
integrate this equation with respect to time, most of the
physically relevant features of the velocity distributions
including large velocity statistics and the time dependent
behavior of the moments can be found analytically, as will be
shown below.

\section{Scaling Solutions}

Numerical simulations in two-dimensions suggest that the velocity
distribution approaches the scaling form \cite{bmp}
\begin{equation}
\label{pscl} P({\bf v},t)\sim
{1\over T^{d/2}}{\cal P}\left({v\over \sqrt{T}}\right).
\end{equation}
The scaling form of the Fourier transform reads
\begin{equation}
\label{Fscal} F({\bf k},t)=\Phi(x),\qquad {\rm with} \qquad
x=k^2T.
\end{equation}
In the $k\to 0$ limit, the Fourier transform behaves as $F({\bf
k},t)\cong 1-{1\over 2}\,k^2\, T$.  This implies that the first two
terms in the Taylor expansion of the corresponding scaling function
are universal, $\Phi(x)\cong 1-{1\over 2}x$. Substituting the above
scaling form into the governing equation (\ref{fkt1}) and using the
temperature cooling rate ${d\over d\tau}T=-\lambda T$ yields the
governing equation for the scaling function
\begin{equation}
\label{Fx} -\lambda x\,\Phi'(x)+\Phi(x) =\big\langle \Phi(\xi
x)\,\Phi(\eta x)\big\rangle.
\end{equation}
One can check that the velocity distribution is purely Maxwellian
$\Phi(x)=e^{-x/2}$ in the elastic case \cite{caveat}. Indeed, $\lambda=0$ and 
$\xi+\eta=1$ in this case. A stochastic process of elastic collisions
effectively randomizes the velocities and leads to a thermal
distribution.

\subsection{Algebraic tails}

It is instructive to consider first the one-dimensional case.  Here,
integration over $\mu$ is immediate as this variable equals unity, and the
scaling function satisfies $-\lambda x\,\Phi'(x)+\Phi(x) 
=\Phi\left[\epsilon^2 x\right]\,\Phi\left[(1-\epsilon)^2 x\right]$.
Remarkably, this non-local non-linear differential equation admits a very
simple solution \cite{bmp}
\begin{equation}
\label{F1sol}
\Phi(x)=\left(1+\sqrt{x}\,\right)\,e^{-\sqrt{x}}.
\end{equation}
Performing the inverse Fourier transform gives the velocity
distribution as a squared Lorentzian
\begin{equation}
\label{Pvt}
{\cal P}(w)={2\over \pi}\left(1+w^2\right)^{-2}.
\end{equation}
The scaling solution (\ref{Pvt}) is universal as it is independent
of the dissipation coefficient $\epsilon$. Its key feature is the
algebraic tail, ${\cal P}(w)\sim w^{-4}$ as $|w|\to\infty$.

In general dimension $d$, the large velocity behavior of the velocity
distribution can be determined from the small wave number behavior of
its Fourier transform. For example, the small-$x$ expansion of the
one-dimensional solution (\ref{F1sol}) contains both regular and
singular terms: \hbox{$\Phi(x)=1-{1\over 2}\,x+{1\over
3}\,x^{3/2}+\cdots$}, and the dominant singular $x^{3/2}$ term
reflects the $w^{-4}$ tail of ${\cal P}(w)$.  In general, an algebraic
tail of the velocity distribution (\ref{pscl}),
\begin{equation}
\label{tail} {\cal P}(w)\sim w^{-\sigma}\qquad {\rm as}\quad w\to\infty,
\end{equation}
indicates the existence of a singular component in the Fourier transform,
\begin{equation}
\label{sing} \Phi_{\rm sing}(x)\sim x^{(\sigma-d)/2}
\qquad {\rm as}\quad x\to 0.
\end{equation}
The inverse is also correct. This can be seen by recasting the Fourier
transform \hbox{$\Phi(x)\propto \int_0^\infty dw\, w^{d-1}{\cal
P}(w)\,e^{iw\sqrt{x}}$} into a Laplace transform \hbox{$I(s)\propto
\int_0^\infty dw\,w^{d-1}{\cal P}(w)\,e^{-ws}$} by writing
$x=-s^2$. The small-$s$ expansion of $I(s)$ contains regular and
singular components.  For example, when $\sigma<d$, the integral
$I(s)$ diverges as $s\to 0$ and integration over large-$w$ yields the
dominant contribution $I_{\rm sing}(s)\sim s^{\sigma-d}$. When
$d<\sigma<d+1$, $I(0)$ is finite, but the next term is the above
singular term, so \hbox{$I(s)=I(0)+I_{\rm sing}(s)+\cdots$}.  In
general, the singular contribution is $I_{\rm sing}(s)\sim
s^{\sigma-d}$, thereby leading to Eq.~(\ref{sing}).

The exponent $\sigma$ can be now obtained by inserting
$\Phi(x)=\Phi_{\rm  reg}(x)+\Phi_{\rm sing}(x)$ into
Eq.~(\ref{Fx}) and balancing the dominant singular terms. We find
that $\sigma$ is a root of the integral equation
\begin{equation}
\label{main}
1-\lambda\,{\sigma-d\over 2} =\big\langle
\xi^{(\sigma-d)/2}+\eta^{(\sigma-d)/2}\big\rangle.
\end{equation}
This relation, originally derived in Refs.\cite{kb,eb}, can be recast
as an eigenvalue problem.  Indeed, defining $\lambda_{\nu}=\langle
1-\xi^\nu-\eta^\nu\rangle$ we can re-write Eq.~(\ref{main}) as
$\lambda_{\mu}=\mu\lambda_1$ with $\mu={\sigma-d\over 2}$.  Note that
$\lambda\equiv\lambda_1$, so there is an obvious solution $\mu=1$, or
$\sigma=d+2$.  In this case, the singular term simply coincides with
the dominant regular term, $x^{(\sigma-d)/2}=x$. Hence, this solution
is trivial and in the following we shall seek a solution with
$\sigma>d+2$.

The integral equation (\ref{main}) can also be rewritten in terms of
special functions. The first integral on the right-hand side of
(\ref{main}) can be expressed in terms of the hypergeometric function
${}_2F_1(a,b;c;z)$ \cite{aa} and the second as a ratio of beta
functions:
\begin{eqnarray}
\label{solve}
&&1-\epsilon(1-\epsilon)\,{\sigma-d\over d}=
\\
&&{}_2F_1
\left[{d-\sigma\over 2},{1\over2};{d\over 2}; 1-\epsilon^2\right]
+(1-\epsilon)^{\sigma-d}\, {\Gamma\left({\sigma-d+1\over
2}\right)\Gamma\left({d\over 2}\right)\over
\Gamma\left({\sigma\over 2}\right)\,\Gamma\left({1\over
2}\right)}.\nonumber
\end{eqnarray}
We conclude that the exponent $\sigma\equiv \sigma(d,\epsilon)$
depends in a nontrivial fashion on the spatial dimension $d$ as well
as the dissipation coefficient $\epsilon$.

First, let us investigate the dependence on the dissipation coefficient by
considering the quasi-elastic limit $\epsilon\to 0$. In the elastic
case, the Maxwellian distribution, \hbox{$\Phi(x)=e^{-x/2}$,} implies
a diverging exponent $\sigma\to\infty$ as $\epsilon\to 0$.  Therefore,
the right-hand side of Eq.~(\ref{solve}) vanishes in the quasi-elastic
limit and to leading order
\begin{equation}
\label{sig0} 
\sigma\simeq {d\over \epsilon}. 
\end{equation}
Clearly, the quasi-elastic limit is singular. Dissipation, even if
minute, seriously changes the nature of the system
\cite{lcdkg,italy,bcdr}. Further corrections can be obtained via a
systematic perturbation expansion in $\epsilon$. We merely quote the
two leading corrections in the physically relevant dimensions
\begin{eqnarray*}
\sigma(2,\epsilon)&=&{2\over \epsilon}
-{2\left(e^{-2}+1\right)\over\sqrt{\pi\epsilon}}
+{4\pi-(e^{-2}+1)^2\over \pi}+{\cal O}(\epsilon^{1/2}),\\
\sigma(3,\epsilon)&=&{3\over \epsilon}
-\sqrt{3\pi\over 2\epsilon}+\left[6-e^{-3}-{\pi\over 4}\right]
+{\cal O}(\epsilon^{1/2}).
\end{eqnarray*}

Next, we discuss the dependence on the dimension. First, one can
verify that $\sigma=4$ when $d=1$ by utilizing the identity
${}_2F_1(a,b;b;z)=(1-z)^{-a}$. In the infinite dimension limit, the
second integral $\langle \eta^{(\sigma-d)/2}\rangle$ in
Eq.~(\ref{main}) is negligible as it vanishes exponentially with the
dimension. To evaluate the second integral we take the limits
$d\to\infty$ and $\mu\to 0$ with $z=\mu d/2$ fixed.  The integration
measure (\ref{dmu}) is transformed according to ${\cal D}\mu\to (\pi
z)^{-1/2}e^{-z}\,dz$, and the basic Eq.~(\ref{main}) becomes
$1-\epsilon(1-\epsilon) u=\int_0^{\infty} dz\, (\pi z)^{-1/2}\,
e^{-[1+(1-\epsilon^2)u]z}$ with the shorthand notation $u={\sigma\over
d}-1$.  Performing the integration yields $1-\epsilon(1-\epsilon)
u=[1+(1-\epsilon^2)u]^{-1/2}$. This cubic equation has the
aforementioned trivial solution $u=0$ and two non-trivial solutions.
Choosing the physically relevant $u$, we obtain that as $d\to\infty$
\begin{equation}
\label{larged} {\sigma\over d} = {1+{3\over
2}\epsilon-\epsilon^3-\epsilon^{1/2} \left(1+{5\over
4}\epsilon\right)^{1/2}\over \epsilon(1-\epsilon^2)}.
\end{equation}

\begin{figure}
\centerline{\includegraphics[width=8cm]{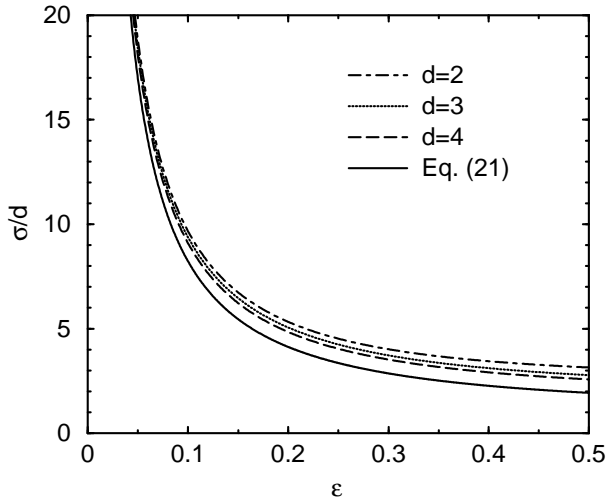}}
\caption{The exact exponent $\sigma$, obtained from Eq.~(\ref{solve}),
versus the dissipation parameter $\epsilon$. The exponent was scaled
by the dimension $d$. Shown also is the limiting large dimension
expression (\ref{larged}).}
\end{figure}

In general, $\sigma\propto d$, and therefore, the algebraic decay
becomes sharper as the dimension increases.  The exponent
$\sigma(d,\epsilon)$ increases monotonically with increasing $d$, and
additionally, it increases monotonically with decreasing $\epsilon$
(see Fig.~1). Both features are intuitive as they mirror the monotonic
dependence of the energy dissipation rate
$\lambda=2\epsilon(1-\epsilon)/d$ on $d$ and $\epsilon$. Hence, the
completely inelastic case provides a lower bound for the exponent,
$\sigma(d,\epsilon)\ge \sigma(d,\epsilon=1/2)$ with
$\sigma(d,1/2)=6.28753$, $8.32937$, for $d=2$, $3$, respectively.
Numerical simulation results are consistent with the former value
\cite{b}.  The algebraic tails are characterized by unusually large
exponents which may be difficult to measure accurately in practice;
for instance, typical granular particles are characterized by the
dissipation coefficient $\epsilon\approx 0.1$ yielding $\sigma\approx
30$ in three dimensions. Figure 1 also shows that the quantity
$\sigma/d$ weakly depends upon the dimension, and the large-$d$ limit
(\ref{larged}) provides a good approximation even at moderate
dimensions.

\subsection{Divergence of the moments}

The algebraic tail of the velocity distribution implies that
sufficiently small moments of the scaling function $\Phi(x)$ are
finite, while moments larger than some index diverge. In the
scaling regime, moments of the velocity distribution can be
calculated by expanding the Fourier transform in powers of $x$,
\begin{equation}
\label{Fexpan} \Phi(x)=\sum_{n\geq 0} \phi_n (-x)^n.
\end{equation}
The coefficients $\phi_n$ yield the leading asymptotic behavior of
the velocity moments, $M_k(t)=\int d{\bf v}\, v^k P({\bf v},t)$, via
the  relation $(2n)!\,T^n\phi_n\simeq \langle \mu^n\rangle
M_{2n}$. Inserting the moment expansion into the governing
equation (\ref{Fx}) yields the closed hierarchy of equations
\begin{eqnarray}
\label{phin}
(\lambda_n-n\lambda_1)\phi_n=\sum_{m=1}^{n-1}
\lambda_{m,n-m} \phi_m\phi_{n-m},
\end{eqnarray}
with $\lambda_n=\langle 1-\xi^n-\eta^n\rangle$ and
$\lambda_{m,l}=\langle\xi^m\eta^l\rangle$. The first few
coefficients are written explicitly in Appendix A. Starting with
$\phi_0=1$ and $\phi_1=1/2$, further coefficients are determined
recursively from (\ref{phin}). In the elastic case ($\epsilon=0$),
one has $\phi_n=(n!\,2^{n})^{-1}$, consistent with $\Phi(x)=e^{-x/2}$. 
For general $\epsilon$, the first two terms
are
\begin{eqnarray}
\label{A2}
\phi_2&=& {1\over 8}\,{1-3\,{1-\epsilon^2\over
d+2}\over 1-3\,{1+\epsilon^2\over d+2}},\\
\label{A3}
\phi_3&=&{1\over 48}\, {1-3\,{1-\epsilon^2\over
d+2}\over 1-3\,{1+\epsilon^2\over d+2}}  \,
{1-3\,{(1-\epsilon)(1+3\epsilon)\over d+2}
+30\,{\epsilon(1-\epsilon)(1-\epsilon^2)\over (d+2)(d+4)}\over
1-3\,{(1+\epsilon)^2\over d+2}
+10\,{\epsilon(1-\epsilon)(3+\epsilon^2)\over
(d+2)(d+4)}}\,.\nonumber
\end{eqnarray}
The behavior is determined by two parameters: $d$ and $\epsilon$.
Fixing $\epsilon$, we see that a given moment $\phi_n$ is finite only
if the dimension is sufficiently large, $d>d_n(\epsilon)$. In
particular, $\phi_n$ is finite only if the left-hand side of
Eq.~(\ref{phin}) is positive, $\lambda_n-n\lambda_1>0$.  This
condition is satisfied only if the dimension is sufficiently large
$d>d_n$, with $d_n$ being the spatial dimension at which
$\lambda_n=n\lambda_1$.  For example, $\phi_2>0$ when $d>d_2$, and
$\phi_3$ is finite only when $d>d_3$ with the following crossover
dimensions
\begin{eqnarray}
\label{dn}
d_2&=&1+3\epsilon^2,\\
d_3&=&{3\over 2}\left(\epsilon^2+2\epsilon-1\right)+{1\over2}
\sqrt{25-60\epsilon+186\epsilon^2-4\epsilon^3+49\epsilon^4}.\nonumber
\end{eqnarray}
Conversely, for a fixed dimension, a given moment is finite only if
the dissipation is sufficiently small.  For example, $\phi_3$ is
positive only when $\epsilon<0.302074,0.427438$ at $d=2,3$. Such
values, obtained by solving polynomial equations yield integer values
of the large velocity decay exponent $\sigma(2,0.302074)=8$ and
$\sigma(3,0.427438)=9$, in accord with direct numerical solution of
Eq.~(\ref{solve}).

\subsection{Velocity correlations}

Maxwell's seminal derivation of the Maxwellian distribution (see
Ref.\cite{classical}, p.~36) relies on two basic assumptions: (1)
Isotropy of the velocity distribution, and (2) Absence of correlations
between the velocity components. The latter assumption is directly
probed using the following correlation measure
\begin{equation}
\label{Q-def} Q= {\langle v_x^2v_y^2\rangle-\langle
v_x^2\rangle\langle v_y^2\rangle \over \langle v_x^2\rangle\langle
v_y^2\rangle}.
\end{equation}
A non-vanishing $Q$ indicates that velocity correlations do exist,
and the larger $Q$ the larger the correlation. In the freely
evolving case, this quantity easily follows from the small-$x$
behavior of the scaling function $\Phi(x)$. By definition,
$\langle v_x^2\rangle=\langle v_y^2\rangle=T$ and furthermore,
$\langle v_x^2v_y^2\rangle={\partial^2\over\partial k_x^2}
{\partial^2\over\partial k_y^2}\,F\Big|_{{\bf
k}=0}=4T^2\Phi''(0)$. Consequently, one has
$Q=4\,\Phi''(0)-1=8\phi_2-1$. Using Eq.~(\ref{A2}) we find
\begin{equation}
\label{Q-sr} Q={6\epsilon^2\over d-(1+3\epsilon^2)},
\end{equation}
when $d>d_2=1+3\epsilon^2$, and $Q=\infty$ otherwise. While the
quantity $Q$ is physical when $d\geq2$, it is sensible to use analytic
continuation to reveal the underlying divergence.  Velocity
correlations vanish for elastic gases. Interestingly, inelasticity
introduces strong velocity correlations, and the larger $\epsilon$ the
larger the correlations as $Q$ increases monotonically with increasing
$\epsilon$. The perfectly inelastic case ($\epsilon=1/2$) again
provides a bound: $Q<Q_{\rm max}=6$, $6/5$ for $d=2$, $3$,
respectively. This behavior is somewhat intuitive as the unisotropic
collision rule (\ref{inel}) discriminates the velocity component
normal to the impact direction.

\section{Nonequilibrium Dynamics}

Thus far, we focused on the leading asymptotic behavior of the
velocity distribution. The diverging moments and the dissipative
nature of this system suggest that the time dependence may exhibit
rich behavior. Thus, we study relaxation of velocity characteristics
such as the moments and the autocorrelation function.

\subsection{Multiscaling of the moments}

While moments of the scaling function diverge, the actual moments must
remain {\em finite} at all times, particularly at the scaling
regime. Therefore, the above moment analysis suggests that knowledge
of the leading asymptotic behavior is not sufficient to characterize
the time dependent behavior of sufficiently large moments.

The time evolution of the moments can be studied using the expansion
\begin{equation}
\label{fullexp}
F(z,\tau)=\sum_{n=0}^\infty f_{n}(\tau)\,(-z)^n.
\end{equation}
The actual moments are related to the coefficients via
$(2n)!f_{n}=\langle \mu^n\rangle M_{2n}$.  Substituting the expansion
(\ref{fullexp}) into (\ref{fkt1}) yields the evolution equations
\begin{equation}
{d\over
d\tau}f_{n}+\lambda_{n}f_{n}=\sum_{m=1}^{n-1}\lambda_{m,n-m}f_{m}f_{n-m}.
\end{equation}

We demonstrate multiscaling asymptotic behavior by evaluating the
second, fourth, and sixth moments. The second moment is obtained from
${d\over d\tau}f_1+\lambda_1 f_1=0$ with
$\lambda_1=\lambda=2\epsilon(1-\epsilon)/d$. Hence, we recover Haff's
law $f_1(\tau)=f_1(0)e^{-\lambda_1\tau}$, or
$f_1(t)=f_1(0)\left(1+t/t_*\right)^{-2}$.  Asymptotically, the second
moment of the velocity distribution has the universal behavior,
$M_2\sim t^{-2}$. The next coefficient $f_2$ satisfies
\begin{equation}
\label{f2} {d\over d\tau}f_2+\lambda_2f_2=\lambda_{1,1}f_1^2.
\end{equation}
Solving Eq.~(\ref{f2}) we find that $f_2(\tau)$ is a linear
combination of two exponentials, $e^{-\lambda_2\tau}$ and
$e^{-2\lambda_1\tau}$, whose decay coefficients are equal
$\lambda_2=2\lambda_1$ at the crossover dimension $d_2$.
Integrating the rate equation (\ref{f2}) and translating back to
the physical time $t$, we obtain
\begin{equation}
\label{f2sola} f_2(t)=C_1\left(1+t/t_*\right)^{-4}+
C_2\left(1+t/t_*\right)^{-2\alpha_2}
\end{equation}
for $d\ne d_2$.  Here, $\alpha_n=\lambda_n/\lambda_1$,
$C_1=\lambda_{1,1}f_1^2(0)/(\lambda_2-2\lambda_1)$,
and $C_2=f_2(0)-C_1$. When $d=d_2$ one finds
\begin{equation}
\label{f2solb} f_2(t)=\big[C_1\ln\left(1+t/t_*\right)+C_2\big]
\left(1+t/t_*\right)^{-4},
\end{equation}
with $C_1=\lambda_{1,1}f_1^2(0)$ and $C_2=f_2(0)$.  Thus for $d>d_2$, the
fourth moment exhibits ordinary scaling, $f_2\sim t^{-4}$, or $M_4\sim
M_2^2\sim t^{-4}$. When $d<d_2$, multiscaling becomes apparent as
$f_2\sim t^{-2\alpha_2}$ and therefore the ratio $M_4/M_2^2$ diverges
asymptotically.  This is consistent with the divergence of the fourth
moment of the scaling function $\Phi(x)$ that occurs at the same
crossover dimension $d_2$. A logarithmic correction occurs at this
dimension. In summary, we find the following leading asymptotic
behavior of the fourth moment 
\begin{equation}
\label{M4all} M_4(t)\sim\cases{t^{-4}&    $d>d_2$,\cr
                        t^{-4}\,\ln t&    $d=d_2$,\cr
                        t^{-2\alpha_2}&    $d<d_2$.}
\end{equation}

A similar calculation can be carried for the sixth moment.  The
solution of \hbox{${d\over d\tau}f_3+\lambda_3f_3=(\lambda_{1,2}+
\lambda_{2,1})f_1f_2$}, with
$f_1$ and $f_2$ given above involves three exponentials:
$e^{-\lambda_3 \tau}$, $e^{-(\lambda_1+\lambda_2)\tau}$, and
$e^{-3\lambda_1\tau}$. Asymptotically, the first exponential dominates
when $d<d_3$, and consequently, $M_6\sim t^{-2\alpha_3}$; otherwise,
the third exponential dominates and thence ordinary scaling occurs,
$M_6\sim t^{-6}$. Generally, the leading asymptotic behavior of the
$2n$-th moment is characterized by two different regimes
\begin{equation}
\label{Mnall} M_{2n}(t)\sim\cases{t^{-2n}&   $d>d_n$,\cr
                           t^{-2n}\,\ln t&   $d=d_n$,\cr
                            t^{-2\alpha_n}&   $d<d_n$.}
\end{equation}
Further logarithmic corrections affecting sub-dominant terms occur at
the crossover dimensions $d_2,\ldots,d_{n-1}$. 

The dependence of $d_n(\epsilon)$ on the dissipation coefficient is
shown in Figure 2. In the physical dimensions $d=2,3$, the fourth
moment exhibits ordinary scaling behavior.  The sixth order moment
exhibits multiscaling if the dissipation coefficient is large enough:
$\epsilon>0.302074, 0.427438$ for $d=2,3$, respectively.  In the large
$n$ limit, $\lambda_n\to 1$ so from $\lambda_n=n\lambda_1$ we find
$d_n\to 2\epsilon(1-\epsilon)\,n$.  Thus, regardless of the
dissipation parameters and the dimension, sufficiently large moments
exhibit multiscaling:
\begin{equation}
\label{multiscaling} 
M_{2n}\propto M_2^{\alpha_n}, \qquad  \alpha_n=\lambda_n/\lambda_1.
\end{equation}
Interestingly, the multiscaling exponents saturate asymptotically,
$\alpha_n\to d/\big[2\epsilon(1-\epsilon)\big]$ as $n\to\infty$.  Of course,
if the dimension increases or the dissipation parameter decreases, the order
of the lowest moment exhibiting multiscaling increases, and in practice, it
may be difficult to observe deviations from ordinary scaling. For example, at
$d=3$ and $\epsilon=0.1$, multiscaling occurs only for moments whose index
exceeds $30$!

\begin{figure}
\centerline{\includegraphics[width=8cm]{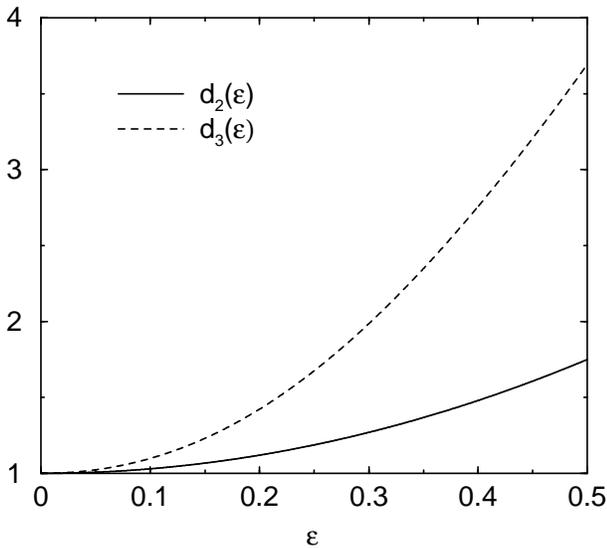}}
\caption{The crossover dimensions $d_n(\epsilon)$ of Eq.~(\ref{dn})
versus the dissipation coefficient for $n=2$, $3$.}
\end{figure}

\subsection{Non-universal velocity autocorrelations}

The autocorrelation function quantifies memory in the velocity of a
tagged particle \cite{classical}. The velocity autocorrelation,
$A(t_w,t)$, is defined via
\begin{equation}
\label{at-def} A(t_w,t)=\overline{{\bf v}(t_w)\cdot{\bf v}(t)}
\end{equation}
where the overline denotes averaging over all particles and $t_w$
is the ``waiting'' time, $t_w<t$.

It is simple to show (see appendix B) that the autocorrelation
evolves according to the following linear equation
\begin{equation}
\label{at-eq}
{d\over d\tau}A(\tau_w,\tau)=-{1-\epsilon\over d}\,A(\tau_w,\tau),
\end{equation}
where time is again expressed in terms of the collision counters $\tau_w$ and
$\tau$. Equation (\ref{at-eq}) is solved to give
$A(\tau_w,\tau)=A(\tau_w,\tau_w)\exp[-{1-\epsilon\over d}(\tau-\tau_w)]$, or
equivalently
\begin{equation}
\label{at}
A(t_w,t)=
A_0\left(1+t_w/t_*\right)^{1/\epsilon-2}\left(1+t/t_*\right)^{-1/\epsilon},
\end{equation}
with $A_0=d\,T_0$. Therefore, $A(t_w,t)$ is a function of the waiting
time $t_w$ and the observation time $t$, and not simply of their
difference, $t-t_w$. This interesting history dependence or ``aging''
is another signature of the nonequilibrium nature of our system.

Memory of the initial conditions can be quantified by setting
$t_w=0$. Writing $A(t)\equiv A(0,t)$ we arrive at the following
algebraic decay
\begin{equation}
\label{a0t} A(t)=A_0\left(1+t/t_*\right)^{-1/\epsilon}.
\end{equation}
In contrast with the temperature which decays with a universal law,
$T(t)\sim t^{-2}$, the autocorrelation decays with a non-universal
law, $A(t)\sim t^{-1/\epsilon}$. The exponent is independent of the
dimension. However, it strongly depends on the dissipation, and the
stronger the dissipation, the stronger the memory of the initial
conditions. This decay exponent is bounded by $2\leq 1/\epsilon\leq
\infty$. In the elastic case, $\epsilon=0$, a simple exponential decay
occurs, and in the totally inelastic case, $\epsilon=1/2$, the
autocorrelation and the temperature are proportional to each other.

The autocorrelation function allows calculation of the long-time
spread in the position of a tagged particle $\Delta^2(t)\equiv\langle
|{\bf x}(t)-{\bf x}(0)|^2\rangle$. Using ${\bf x}(t)-{\bf
x}(0)=\int_0^t dt'{\bf v}(t')$, one can immediately express
$\Delta^2(t)$ via the autocorrelation function, $\Delta^2=2\int_0^t
dt'\int_0^{t'} dt''A(t'',t')$. Substituting (\ref{at}) into this 
expression and performing the integration yields
$\Delta^2(t)=C_1\ln\left(1+t/t_*\right)
+C_2\left[(1+t/t_*)^{1-1/\epsilon}-1\right]$
with $C_1=2A_0t_*^2\epsilon/(1-\epsilon)$ and
$C_2=-C_1\,\epsilon/(1-\epsilon)$.  Asymptotically, the second term is
negligible, and the spread has a generic logarithmic behavior
\begin{equation}
\Delta\sim \sqrt{\ln t}
\end{equation}
reflecting the $t^{-1}$ decay of the overall velocity scale \cite{bp,kh}.

\section{Steady States}

Thus far, we have discussed freely cooling systems where the energy
decreases indefinitely. In typical experimental situations, however,
the system is supplied with energy to balance the energy dissipation
\cite{lcdkg,ou,nbc,rm}.  Theoretically, it is natural to consider
white noise forcing \cite{van,ccg}, i.e., coupling to a thermal heat
bath which leads to a nonequilibrium steady state. Interestingly, a
stretched exponential behavior, $P(v)\propto \exp(-v^{3/2})$, is found
for the driven inelastic hard sphere gas \cite{van}. 

Specifically, we assume that in addition to changes due to collisions,
velocities may also change due to an external forcing:
\hbox{${dv_j\over dt}|_{\rm heat}=\xi_j$} with $j=1,\ldots,d$. We use
standard uncorrelated white noise \hbox{$\langle\xi_i(t)
\xi_j(t')\rangle =2D\delta_{ij}\delta(t-t')$} with a zero average
$\langle \xi_j\rangle=0$.  The rate equation for the temperature is
modified by the additional source term ${d\over dt}T+\lambda
T^{3/2}=2D$, and the system approaches a steady state,
$T_{\infty}=(2D/\lambda)^{2/3}$.  The relaxation toward this state is
exponential, $|T_{\infty}-T|\sim e^{-{\rm const.}\times t}$.

Uncorrelated white noise forcing amounts to diffusion in velocity
space.  Therefore, Eq.~(\ref{fkt}) is modified as follows,
${\partial\over\partial \tau}\to {\partial\over\partial
\tau}+Dk^2$. In the steady state, the Fourier transform,
$F_\infty({\bf k})\equiv \Psi(y)$ with $y=Dk^2$, obeys
\begin{equation}
\label{int} (1+y)\,\Psi(y)=\big\langle \Psi(\xi y)\, \Psi(\eta
y)\big\rangle.
\end{equation}
This equation is solved recursively by employing the cumulant expansion
\begin{equation}
\Psi(y)=\exp\left[\sum_{n=1} \psi_n(-y)^n\right].
\end{equation}
The cumulants $\kappa_n$, defined as
\begin{equation}
F_\infty({\bf k})=\exp\left[\sum_{m=1}^\infty {\kappa_m(ik)^m\over
m!}\right],
\end{equation}
are related to the coefficients $\psi_n$, viz. $\kappa_n=(2n)!D^n\psi_n$.
Writing $1+y=\exp\left[\sum_{n\geq 1}(-y)^n/n\right]$, we recast
Eq.~(\ref{int}) into 
\begin{eqnarray}
\label{cumeqg} 1=\Bigg\langle\exp\left[-\sum_{n=1}^{\infty}
\left(\widetilde \psi_n-n^{-1}\right)(-y)^n\right]\Bigg\rangle,
\end{eqnarray}
with the auxiliary variables
$\widetilde\psi_n=\psi_n(1-\xi^n-\eta^n)$.  The desired cumulants
$\psi_n$ are obtained by evaluating recursively the angular integrals
of the auxiliary variables, $\langle \widetilde\psi_n\rangle$, and
then using the identities $\psi_n=\langle
\widetilde\psi_n\rangle/\lambda_n$.  In one dimension, $\langle
\mu^n\rangle=1$ and one immediately obtains $\langle
\widetilde\psi_n\rangle=n^{-1}$, and consequently
$n\psi_n=[1-\epsilon^{2n}-(1-\epsilon)^{2n}]^{-1}$\cite{bk}.  In
higher dimensions, the quantities $\langle \widetilde\psi_n\rangle$
acquire non-trivial dependence on $n$, e.g., $\langle
\widetilde\psi_1\rangle =1$, $\langle \widetilde\psi_2\rangle={1\over
2}\langle \widetilde\psi_1^2\rangle$, and $\langle
\widetilde\psi_3\rangle=\langle
\widetilde\psi_1\widetilde\psi_2\rangle-{1\over 6}\langle
\widetilde\psi_1^3\rangle$. The first few values for $\psi_n$ can be
then evaluated.  In particular, $\psi_1=1/\lambda_1$ and
$\psi_2=\big\langle(1-\xi-\eta)^2\big\rangle/\big(2\lambda_1^2\lambda_2\big)$,
from which one can determine explicit expressions:
\begin{eqnarray}
\label{psi1}
\psi_1&=&{d\over 2\epsilon(1-\epsilon)},\\
\label{psi2} \psi_2&=&{3\,d^2\over
4(d+2)(1-\epsilon^2)-12(1-\epsilon)^2(1+\epsilon^2)}.\nonumber
\end{eqnarray}
Thus, the steady state distribution is not purely Maxwellian.

To probe velocity correlations or alternatively, deviations from a
factorizing Maxwellian distribution, we consider the quantity $Q$,
defined in Eq.~(\ref{Q-def}). At the steady state, it is given by
\begin{equation}
\label{Q-def-std}
Q={\Psi''(0)\over [\Psi'(0)]^2}-1.
\end{equation}
In terms of the first two coefficients of the cumulant expansion,
$Q=2\psi_2/\psi_1^2$. Substituting the value of these coefficients
yields
\begin{equation}
\label{Q-std}
Q={6\epsilon^2(1-\epsilon)\over
(d+2)(1+\epsilon)-3(1-\epsilon)(1+\epsilon^2)}.
\end{equation}
Note that for a fixed spatial dimension, this quantity is maximal in
the completely inelastic case. For instance, $Q_{\rm max}=2/11$ in two
dimensions and $Q_{\rm max}=2/15$ in three dimensions.  These values
are smaller by an order of magnitude or more than the corresponding
values in the unforced case. Intuitively, one expects that white noise
forcing randomizes the velocities of the particles. Indeed, velocity
correlations are much less pronounced in this case, as seen in Figure
3.  Additionally, velocity correlations diminish as the dimension
increases. At large dimensions, velocity correlations vanish according
to $Q\sim d^{-1}$, indicating that the velocity distribution becomes
purely Maxwellian, $\Psi(y)\to \exp(-y/2)$, when $d\to\infty$.

\begin{figure}
\centerline{\includegraphics[width=8cm]{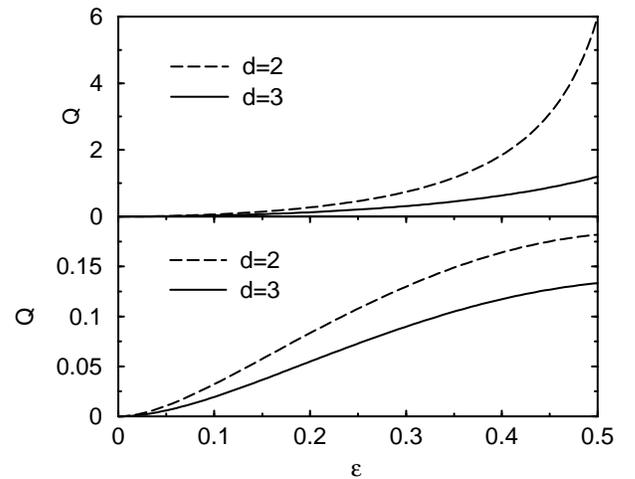}}
\caption{The velocity correlation measure $Q$ versus the dissipation
coefficient $\epsilon$. The scaling regime result (\ref{Q-sr}) is
shown in the top graph, and the steady state result (\ref{Q-std}) is
shown in the bottom graph.}
\end{figure}

\section{Generalizations}

The above results can be generalized in a number of ways. For example,
the development of spatial correlations can be considered by placing
particles on a lattice and allowing for nearest-neighbor collisions
only.  In this section we briefly mention two straightforward
generalizations to: (1) energy generating collisions, and (2)
distribution of restitution coefficients.

Thus far, we discussed only the physical case of dissipative
collisions, namely $\epsilon<0$. However, the above results in the
freely evolving case hold for energy generating collisions, i.e.,
$\epsilon>0$ as well. Although the typical velocity scale diverges,
the velocity distribution still follows the scaling solution
(\ref{pscl}) with algebraic large velocity statistics.  The
corresponding exponent $\sigma$ is still obtained from
Eq.~(\ref{solve}).  However, the behavior does change, as follows from
the analytically tractable $d\to\infty$ behavior. In contrast with the
dissipative case, the second term on the right-hand side of
Eq.~(\ref{solve}) now dominates, and it grows exponentially as
$\propto a^{(\sigma-d)/2}$.  Since the left-hand side of
Eq.~(\ref{solve}) is of order unity, the constant $a$ must be equal to
one.  On the other hand, the constant $a$ is evaluated using the
Stirling formula $\Gamma(x)\sim (x/e)^ x$ to give
\hbox{$a=(1-\epsilon)^2 (\sigma-d) d^{d/(\sigma-d)}
\sigma^{-\sigma/(\sigma-d)}$}.  Equating $a=1$, we arrive at
\begin{equation}
\label{larged2} \sigma\cong d\, \nu, \qquad{\rm with} \qquad
(\nu-1)\,\nu^{-{\nu\over \nu-1}}=(1-\epsilon)^{-2}.
\end{equation}
While the exponent rises linearly with the dimension, it exhibit
different $\epsilon$-dependence. Numerical solution of $\nu$ shows
that this large dimension estimate again yields a useful approximation
even at moderate dimensions.

Several recent studies have used a distribution of restitution
coefficients to model driven granular systems, including for example,
a one-dimensional gas of rods with internal degrees of freedom
\cite{gz1,az}, and vertically vibrated layers \cite{btf}.  By tuning
the distribution properly, one can have a situation were overall,
energy is conserved as dissipative collisions are balanced by energy
generating collisions. When the restitution coefficient is drawn from
the distribution $\rho(\epsilon)$, one simply integrates the collision
integral in the Boltzmann equation (\ref{be}) with respect to the
measure $\rho(\epsilon)$. In one dimension, one can check that the
scaling solution $\Phi(x)=\big(1+\sqrt{x}\big)e^{-\sqrt{x}}$ still
holds, and in particular the exponent $\sigma=4$ is robust.  In
general dimension, the exponent $\sigma$ is given by
\begin{equation}
1-\lambda\,{\sigma-d\over 2} =\int d\epsilon\,\rho(\epsilon)
\big\langle \xi^{(\sigma-d)/2}+\eta^{(\sigma-d)/2}\big\rangle
\end{equation}
with the decay rate $\lambda=\int d\epsilon
\rho(\epsilon)\lambda(\epsilon)$. We conclude that algebraic
large-velocity statistics extend to situations where the dissipation
coefficient $\epsilon$ is drawn from a given distribution.

\section{Conclusions}

We have studied inelastic gases within the framework of the Maxwell model, a
Boltzmann equation with a uniform collision rate.  We have shown that this
kinetic theory is analytically tractable as closed evolution equations
characterize the Fourier transform and consequently moments of the velocity
distribution.  In the freely evolving case, the system approaches a scaling
regime, and the velocity distribution has an algebraic large velocity tail.
The corresponding exponent varies continuously with the spatial dimension and
the degree of dissipation.  The decay exponents can be very large and
therefore it may be difficult to distinguish a power law from a stretched
exponential.  In the driven case, we have determined the cumulants of the
velocity distribution.

The time dependent behavior displays a number of interesting
features. Moments of the velocity distribution exhibit multiscaling
asymptotic behavior, and knowledge of the typical velocity is
insufficient to characterize all moments. The velocity autocorrelation
decays algebraically with time, and the corresponding exponent depends
on the restitution coefficient only. 

In contrast with elastic collisions, stochastic inelastic collision processes
are not effective in mixing particle velocities. The stronger the
inelasticity, the stronger the history dependence, i.e., memory of previous
behavior.  Additionally, inelasticity can generate significant correlations
between different velocity components.  Such correlations do develop even in
the forced case, where dissipation is balanced by energy input, and one may
expect that Maxwellian velocity distribution emerge.

The Maxwell model is truly mean field in nature with all aspects of
the collision process being random. While it is not surprising that
such a theory is solvable, the rich structure of the solution is
somewhat unexpected.  For example, the exponent follows from a
transcendental equation, and can not be obtained from heuristic
arguments or dimensional analysis. Remarkably, even the leading
asymptotic behavior in the large dimension limit remains nontrivial as
it involves roots of cubic or transcendental equations.

We have explored only the basic characteristics. Clearly, one can
study higher order velocity correlation measures as well as higher
order autocorrelations.  Furthermore, the relaxation toward the steady
state appears analytically tractable.  The straightforward analysis is 
cumbersome and it may be useful to expand first the solutions in terms
of more natural building blocks, e.g., orthogonal polynomials.

We stress that the Maxwell model is exact for stochastic inelastic
collision processes with random collision partners and impact angles.
It may be applicable in situations where an effective stirring
mechanism leads to perfect mixing. Otherwise, it should be regarded as
an uncontrolled approximation of the Boltzmann equation.  Indeed,
existing theoretical and numerical studies give little evidence for
algebraic tails characterizing inelastic gases.  The only exception
was observed in a system with random restitution coefficients drawn
from a broad distribution. In one dimension, both molecular dynamics
simulation and direct integration of the Boltzmann equation for
inelastic hard spheres show that the velocity distribution has a power
law tail \cite{btf}.

In conclusion, our results, combined with previous kinetic theory studies
that find exponential, stretched exponential, and Gaussian tails,
indicate that extremal velocity characteristics can be sensitive to
the details of the model, let alone parameters such as the restitution
coefficient, and the dimension.

\bigskip\noindent 

We thank A.~Baldassari and M.~H.~Ernst for fruitful correspondence,
and H.~A.~Rose for useful discussions.  This research was supported by
DOE (W-7405-ENG-36), NSF(DMR9978902), and ARO (DAAD19-99-1-0173).

\appendix
\section{The $\lambda$-coefficients}

To compute the coefficients $\lambda_n=\langle
1-\xi^n-\eta^n\rangle$ and $\lambda_{n,m}=\langle\xi^n\eta^m\rangle$ we use
$\xi=1-(1-\epsilon^2)\mu$  and $\eta=(1-\epsilon)^2\mu$. Thus, the following 
integrals are required 
\begin{eqnarray*}
\langle\mu^n\rangle= {\Gamma\left({d \over 
2}\right)\,\Gamma\left(n+{1\over 2}\right) \over \Gamma\left({1\over
2}\right)\,\Gamma\left(n+{d \over 2}\right)}= {1\over d}{3\over
2+d}\cdots {2(n-1)+1\over 2(n-1)+d}.
\end{eqnarray*}
In particular, $\langle \mu\rangle = {1\over d}, \langle \mu^2\rangle={3\over
  d(d+2)}, \langle \mu^3\rangle ={15\over d(d+2)(d+4)}$, so the first few
coefficients are
\begin{eqnarray*}
\lambda_1   &=&2\epsilon(1-\epsilon)\,{1\over d},\\
\lambda_2   &=&2(1-\epsilon^2)\,{1\over d}
              -2(1-\epsilon)^2\left(1+\epsilon^2\right){3\over d(d+2)},\\
\lambda_3 &=& 3(1-\epsilon^2)\,{1\over d}-3(1-\epsilon^2)^2\,{3\over d(d+2)} \\
&&+2\epsilon(1-\epsilon)^3(3+\epsilon^2){15\over d(d+2)(d+4)},\\
\lambda_{1,1}&=&(1-\epsilon)^2\,{1\over d}
               -(1-\epsilon)^2(1-\epsilon^2)\,{3\over d(d+2)}.
\end{eqnarray*}

\section{The autocorrelation evolution equation}

It is useful to work with the collision counter $\tau$. In an
infinitesimal time interval $\Delta\tau$, the velocity of a particle
changes from ${\bf v}\equiv {\bf v}(\tau)$ to
\begin{eqnarray*}
{\bf v}(\tau+\Delta\tau)= \cases{ {\bf v} &prob.
$1-\Delta\tau$,\cr {\bf v}-(1-\epsilon)({\bf v}-{\bf u})\cdot
{\bf n}\, {\bf n}         &prob. $\Delta\tau$.}\nonumber
\end{eqnarray*}
Here ${\bf u}$ is chosen randomly from all particles and the
impact direction ${\bf n}$ is drawn from a uniform distribution.
The rate of change in the autocorrelation function
$A(\tau_w,\tau)=\overline {{\bf v}(\tau_w)\cdot{\bf v(\tau)}}$ is
evaluated as follows
\begin{eqnarray}
\label{integration}
{d\over d\tau}&A&(\tau_w,\tau)=
\lim_{\Delta\tau\to 0}\overline{{\bf v}(\tau_w)\cdot
\left[{\bf v}(\tau+\Delta\tau)-{\bf v}(\tau)\right]/ \Delta\tau}\nonumber\\
=&-&(1-\epsilon)\int d{\bf u}\, P({\bf u},\tau)\int d{\bf n}\,
\overline{\left[\,{\bf v}(\tau_w)\cdot {\bf n}\right]\,
\left[({\bf v}-{\bf u})\cdot {\bf n}\right]}\nonumber\\
=&-&{1-\epsilon\over d}\,\overline{{\bf v}(\tau_w)\cdot{\bf
v}(\tau)} +{1-\epsilon\over d}\,\int d{\bf u}\,P({\bf u},\tau)\,
\overline{{\bf v}(\tau_w)\cdot{\bf u}}\nonumber\\
=&-&{1-\epsilon\over d}A(\tau_w,\tau).
\end{eqnarray}
The angular integration in the second line of Eq.~(\ref{integration})
was performed using the identity
\begin{equation}
\label{id}
H({\bf a},{\bf b})
=\int d{\bf n} \,\,({\bf a}\cdot {\bf n})\,\,({\bf b}\cdot {\bf n})
={1\over d}\,({\bf a}\cdot {\bf b}).
\end{equation}
This identity can be deduced by re-writing the integral as $H({\bf a},{\bf
  b})={\bf a}\cdot {\bf h}({\bf b})$. By symmetry, ${\bf h}({\bf b})=\int
d{\bf n} \,{\bf n}\,({\bf b}\cdot {\bf n})$ is a vector along ${\bf b}$, say
$\Lambda{\bf b}$, implying $H({\bf a},{\bf b})=\Lambda\,({\bf a}\cdot {\bf
  b})$.  Evaluating the special case $H({\bf a},{\bf a})=\langle \mu \rangle\,
a^2$ we obtain (\ref{id}).  Finally, the second term in the third line
vanishes, $\overline{{\bf v}(\tau_w)\cdot{\bf u}(\tau)}=0$, since the
velocity ${\bf u}(\tau)$ of the randomly chosen collision partner is
uncorrelated with ${\bf v}(\tau_w)$.

\end{multicols}

\begin{thebibliography}{99}
\bibitem{caveat}
      Every initial velocity distribution evolves towards the
      Maxwellian distribution if $d>1$ and $\epsilon=0$; for $d=1$,
      there is no relaxation in the elastic case.
\end{thebibliography}
\end{document}